\documentclass[12pt,preprint2]{aastex}

\usepackage{graphicx}

\def \aap {A\&A}

\newcommand{\citeN}[1]{\citeauthor{#1} (\citeyear{#1})}
\newcommand{\citeNP}[1]{\citeauthor{#1} \citeyear{#1}}

\shortauthors{Socas-Navarro \& Trujillo Bueno}
\shorttitle{The Polarization Profiles of \ion{He}{1} 10830}

\begin{document}

\title{Signatures of Incomplete Paschen-Back Splitting in the 
Polarization Profiles of the \ion{He}{1} 10830 \AA \, multiplet}

\author{H. Socas-Navarro}
   	\affil{High Altitude Observatory, NCAR\thanks{The National Center
	for Atmospheric Research (NCAR) is sponsored by the National Science
	Foundation.}, 3450 Mitchell Lane, Boulder, CO 80307-3000, USA}
	\email{navarro@ucar.edu}

\author{J. Trujillo Bueno	\thanks{Consejo Superior de Investigaciones
    Cient\' \i ficas, Spain} 
} 
   	\affil{Instituto de Astrof\'\i sica de Canarias, Avda V\'\i a L\'
   	actea S/N, La Laguna 38205, Tenerife, Spain}
	\email{jtb@iac.es}

\author{E. Landi Degl'Innocenti}
        \affil{Dipartimento di Astronomia e Scienza dello Spazio, Largo
        E. Fermi 2, 50125 Firenze, Italy}
	\email{landie@arcetri.astro.it}

\date{}%

\begin{abstract}
We investigate the formation of polarization profiles induced by a magnetic
field in the \ion{He}{1} multiplet at 10830~\AA . Our analysis considers
the Zeeman splitting in the incomplete Paschen-Back regime. The
effects turn out to be important and produce measurable signatures on the
profiles, even for fields significantly weaker than the level-crossing
field ($\sim$400~G). When compared to profiles 
calculated with the usual linear Zeeman effect, the incomplete Paschen-Back
profiles exhibit the 
following conspicuous differences: a)~a non-Doppler blueshift of the Stokes~V
zero-crossing wavelength of the blue component; b)~area and peak asymmetries,
even in the absence of velocity and magnetic gradients; c)~a $\sim$25\%
reduction in the amplitude of the red component. These features do not vanish
in the weak field limit. The spectral signatures that we analyze in
this paper may be found in previous observations published in the literature.
\end{abstract}
   
\keywords{line: profiles -- 
           Sun: atmosphere --
           Sun: magnetic fields --
           Sun: chromosphere}

\section{Introduction}
\label{intro}

Many of the physical challenges of solar and stellar physics arise from
magnetic processes taking place in the outer regions of the atmosphere. 
Unfortunately, our present empirical knowledge
on the magnetism of the chromosphere,
transition region and corona is still very 
crude. In our opinion,
there are two key avenues that should be pursued to improve the situation.

Firstly, new instrumentation is needed, including 
a space-borne UV polarimeter and a large
ground-based solar telescope optimized for (spectro-)polarimetric
observations (e.g., the Advanced Technology Solar Telescope).
Secondly, but equally importantly, we need appropriate diagnostic tools that
will permit us to infer the magnetic field vector
from the observed polarization in suitably chosen spectral (chromospheric and
coronal) lines. The present paper represents a contribution to this second
line of research. 

In particular, our aim here is to demonstrate that the
determination of the magnetic field vector via the analysis
of the observed polarization in the \ion{He}{1} 10830~\AA \, multiplet
must be carried out considering the wavelength positions and the strengths
of the Zeeman components in the incomplete
Paschen-Back effect regime. Previous research done with this multiplet has
ignored this effect (with the only exception of \citeNP{TBLdIC+02}), which
resulted in considerable confusion surrounding the formation physics of these
lines.

The \ion{He}{1} 10830~\AA \, multiplet originates between a lower
term ($2^3{\rm S}_1$) and an upper term ($2^3{\rm P}_{2,1,0}$).
Therefore, it comprises three spectral lines (see, e.g., \citeNP{RS85}):
a `blue' component at 10829.09~\AA\ with $J_l=1$ and $J_u=0$ (hereafter
Transition~1, or Tr1 for abbreviation),
and two `red' components at 10830.25~\AA\ with $J_u=1$
(hereafter, Tr2) and at 10830.34~\AA\ 
with $J_u=2$ (hereafter, Tr3)  which appear 
blended at solar atmospheric temperatures.
As shown in Fig~\ref{energy}, the energy difference between the levels with
$J_u=2$ and $J_u=1$ is such that their respective magnetic
substates cross each other for magnetic strengths between 400~G and 1500~G,
approximately. Therefore, {\em at least} in this range of strengths,
the Zeeman splitting of such $J$-levels is comparable to their energy separation.
As a result, the perturbation theory of the familiar Zeeman effect is no
longer valid. The energy positions of the magnetic substates as a function of
the field strength have to be calculated in the incomplete Paschen-Back
effect regime, which requires to diagonalize the {\em total} Hamiltonian
(see, e.g., \citeNP{CS70}). Interestingly, as we shall see below, the
proper calculation of the Zeeman 
components results in significant differences in the Stokes profiles 
that emerge from a magnetized stellar atmosphere. Furthermore, such
differences do not vanish in the limit of weak fields.

The outline of this paper is as follows. Section 2 presents a brief
review of the Paschen-Back effect theory,
illustrating the range of magnetic strengths for which the
$J$-levels of the $2^3{\rm P}_{2,1,0}$ term cross. Section
3 shows a detailed comparison of the results that we have obtained
assuming linear Zeeman splitting (LZS) or incomplete Paschen-Back 
splittings (IPBS), including an analytical demonstration that explains
our numerical radiative transfer results in a Milne-Eddington model. Finally,
Section 4 gives our concluding remarks. 

\section{The Paschen-Back effect}
\label{ipb}

According to the theory of the Zeeman effect (see, e.g., \citeNP{CS70}) the
corrections to the degenerate energy $E_J$ of any atomic level of total
angular momentum quantum number $J$ are obtained by diagonalizing the
matrix $<J M | H_B | J M'>$, where $H_B$ is the magnetic Hamiltonian, given
by:
\begin{equation}
H_B=\mu_0 ( \vec L + 2 \vec S ) \vec B \, ,
\end{equation}
with $\mu_0=9.27 \times 10^{-21}$~erg~G$^{-1}$ being the Bohr magneton. By
choosing the 
quantization z-axis of total angular momentum along the magnetic field
vector, one finds:
\begin{equation}
<J M | H_B | J M'> = \mu_0 B g M \delta_{M M'} \,
\end{equation}
where $g$ is the Land\'e factor. This equation shows that, to first order
of perturbation theory, any atomic level of total angular momentum quantum
number $J$ is split by the action of a magnetic field into $(2J+1)$ {\em
  equally spaced} sublevels, the splitting being proportional to the Land\'e
factor $g$ and to the magnetic field strength. It is important to note that a
state of the form $|J M >$ is an eigenvector of the {\em total} Hamiltonian
$(H_0 + H_B)$, with eigenvalue $E_J+\mu_0 B g M$, only when the z-axis of the
reference system points in the direction of the magnetic field. 

The previous well-known results of first-order perturbation theory are
correct only if the splitting produced by the magnetic field on a $J-$level
is small compared to the energy separation between the different $J-$levels
of the $(L,S)$ term under consideration. In other words, the standard theory
of the Zeeman effect is valid only in the limit of ``weak'' magnetic
fields. Here, ``weak'' means that the coupling of either the spin or the
orbital angular momentum to the magnetic field is {\em weaker} than the
coupling between the spin and the orbital angular momentum (the spin-orbit
coupling). This 
is the so-called {\em Zeeman effect regime}.

In the opposite limit, the magnetic field is so ``strong'' that the
spin-orbit interaction can be considered as a perturbation compared to the
magnetic interaction. In this case the magnetic field dissolves the fine
structure coupling -- that is, $\vec L$ and $\vec S$ are practically
decoupled and precess independently around $\vec B$. Therefore, the quantum
number $J$ loses its meaning. In this so-called {\em complete Paschen-Back
  effect regime} the magnetic Hamiltonian is diagonal on the basis $|L S M_L
M_S>$, and the eigenvalues are given by:
\begin{equation}
<L S M_L M_S | H_B | L S M_L M_S > = \mu_0 B (M_L + 2 M_S) \, .
\end{equation}
In this ``strong'' field regime, the term $(L,S)$ splits into a number of
components, each of which corresponds to particular values of $(M_L + 2
M_S)$. In the presence of a magnetic field, the corrections to the energy
of the state $|L S M_L M_S>$ due to the 
spin-orbit interaction has the form $<A \vec L \cdot \vec S>=A M_L M_S$, where
$A$ is a constant with dimensions of energy.

\begin{figure}
\plotone{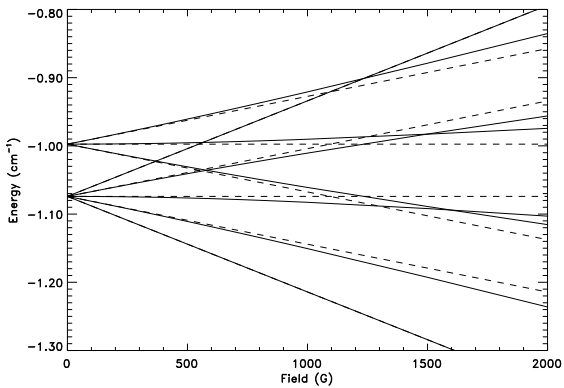}
\caption{Energy diagram for the upper levels of Tr2 and Tr3 (the
  $^3P_1$ and $^3P_2$ levels of the multiplet). Solid: IPBS. Dashed:
  LZS. Note the numerous level crossings between 400 and 1500~G,
  approximately. Energies are in cm$^{-1}$ with respect to the J=0 level.
\label{energy}
}
\end{figure}

It is of interest to note that, since the spin-orbit coupling increases
rapidly with increasing nuclear charge, the conditions for a ``strong'' field
are met at a much lower field with light atoms (like helium) than with heavy
atoms. In fact, as shown in Fig~\ref{energy}, the levels with $J=2$ and $J=1$
of the upper term $^3P$ of the \ion{He}{1} 10830~\AA \, multiplet cross for
magnetic strengths between 400~G and 1500~G, approximately. This
level-crossing regime corresponds to the {\em incomplete Paschen-Back effect
  regime}, in which the energy eigenvectors are gradually evolving from the
form $|L S J M >$ to the form $|L S M_L M_S >$ as the magnetic field
increases. This range between the limiting cases of ``weak'' fields
(Zeeman effect regime) and ``strong'' fields (complete Paschen-Back regime)
is more difficult to analyze since it implies the evaluation of matrix
elements of the form $< J M | H_B | J' M' >$, which (in general) can only be
done numerically. To this aim, we have developed a numerical code capable of
computing, for a magnetic field of arbitrary intensity, the strenghts and the
splittings of the $\sigma$ and $\pi$ components in a line multiplet arising
from the transition between two $(L,S)$ terms. This code is very similar to a
previous one that solves the problem of the intermediate Paschen-Back
effect in a line with hyperfine structure (\citeNP{LdI78}).

\section{Results}
\label{results}

We have developed a numerical code for the synthesis and inversion of Stokes
profiles in a Milne-Eddington atmosphere, considering IPBS as
described in \S\ref{ipb}. This code has been built with the aim of analyzing
spectropolarimetric observations of solar active regions, but we shall use it
here to explore the effects of IPBS on the 10830 polarization profiles.  

\begin{figure*}
\epsscale{1.6}
\plotone{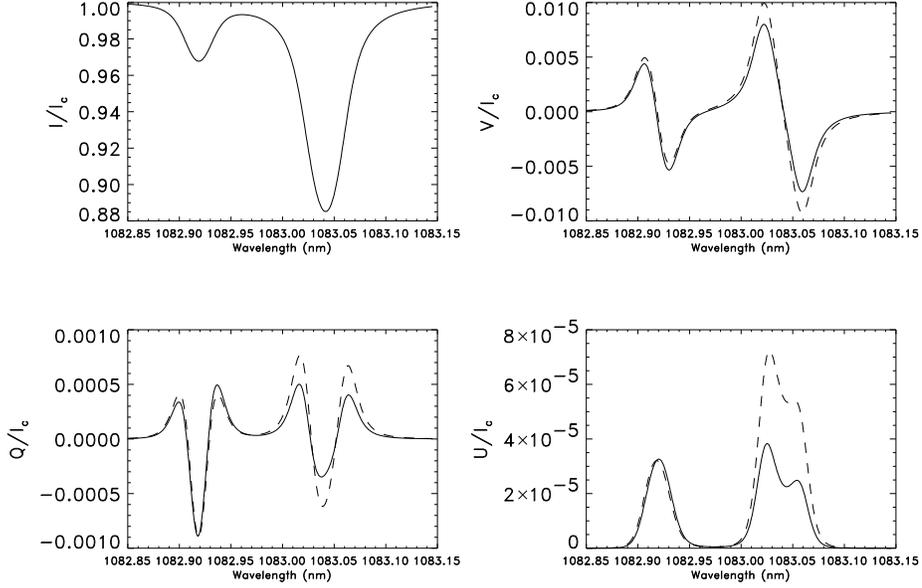}
\caption{Simulated Stokes profiles of the 10830~\AA \,  
  multiplet, assuming a 500~G field inclined 50$^{\rm o}$ with respect to the
  line of sight. The profiles are normalized to the quiet sun continuum
  intensity ($I_c$). 
\label{fig:profiles}
}
\end{figure*}

We consider a reference model atmosphere and calculate the multiplet
component positions and strengths for various magnetic fields (ranging
from 0 to 3~kG). The Zeeman components are
calculated accounting for IPBS and then compared to the usual approximation
of LZS. The most relevant parameters of the
chosen Milne-Eddington model are
$\eta_0=4$, $\Delta \lambda_D=105$~m\AA , $B_0=0.8$, $B_1=0.2$, $a=0.8$
(see, e.g., 
\citeNP{LdI92} for definitions of the Milne-Eddington parameters). Sample
profiles are depicted in Fig~\ref{fig:profiles} for a 500~G field with an
inclination of 50$^{\rm o}$ from the line of sight. There are a few
obvious differences between the IPBS and LZS Stokes~V profiles: a)~The blue
component of the multiplet, Tr1, shows a net asymmetry with a global offset
towards negative values;\footnote{
The sign of this offset depends on the polarity of the profile. It is
produced by an 
imbalance in the weights of the $\sigma$ components of this transition, with
the red component being stronger.}
b)~A small non-Doppler blueshift of the same component (which is actually
related to 
the asymmetry); c)~A reduced amplitude of the red and blue lobes of the red
component of the multiplet 
(Tr2 and Tr3). The linear polarization profiles exhibit also some small
asymmetry. These differences are due to IPBS effects on the Zeeman
components of the transitions, discussed in \S\ref{comp} below.

\begin{deluxetable}{cccccc}
  \tablewidth{0pt}
  \tablecaption{Atomic parameters used in the calculations
  \label{TblAtom}}
  \tablehead{$\lambda$ (\AA ) & Lower term & Upper term & $g_{eff}$ &
    Relative strength}
  \startdata
  10829.09 & $^3$S$_1$ & $^3$P$_0$ & 2.00 & 0.111 \\
  10830.25 & $^3$S$_1$ & $^3$P$_1$ & 1.75 & 0.333 \\
  10830.34 & $^3$S$_1$ & $^3$P$_2$ & 1.25 & 0.556 \\
 \enddata
\end{deluxetable}

In order to compare the IPB results with the LZS ones, one needs the relative
strengths and effective Land\'e factors of the three components. These are
listed in
Table~\ref{TblAtom}. Note that some of the values do not coincide with
those used by other authors in previous works. For example, 
concerning the LZS calculations the effective
Land\'e factor ($g_{eff}$)
of Tr3 is 1.25, whereas \citeN{RSL95} and \citeN{LWK+04} take it to be
0.875. The relative strengths\footnote{
Note that the symbol $f_i$, which is used here and in other previous works on
this multiplet, refers to the relative strengths of the components. These are
not the ``oscillator strengths'' of the transitions, which are also commonly
denoted by $f_i$.
} ($f_i$) of Tr1, Tr2 and Tr3 are here 0.111,
0.333 and 
0.556, respectively. Our $f_i$ agree with those of \citeN{RSL95}, but not
with \citeN{LWK+04} (these authors quote {\em relative oscillator strengths},
but the 
conversion is straightforward). It is not completely clear to us what the
source of the 
discrepancy is. For the $g_{eff}$ we used its standard definition (e.g.,
\citeNP{LdI92}) applied to the quantum numbers shown in the table. The $f_i$
are calculated according to the usual formulae (e.g., \citeNP{RS85}).
Thus, both of these quantities are
derived straightforwardly from well-known atomic parameters. As we discuss
below, it is not possible to reproduce the observations using the LZS
approximation. It is conceivable that other authors had to modify
their atomic parameters 
in an attempt to compensate for IPBS effects. This would allow them to fit 
observed profiles with those calculated using LZS. For example, the ratio of
Stokes~V amplitudes of the blue and red components calculated with LZS is
significantly larger than it should be. Artificially reducing the $g_{eff}$
of Tr3 reduces the amplitude of the red component without affecting the 
intensity profiles, thus bringing the profiles closer to the observations.

\begin{figure*}
\epsscale{1.6}
\plotone{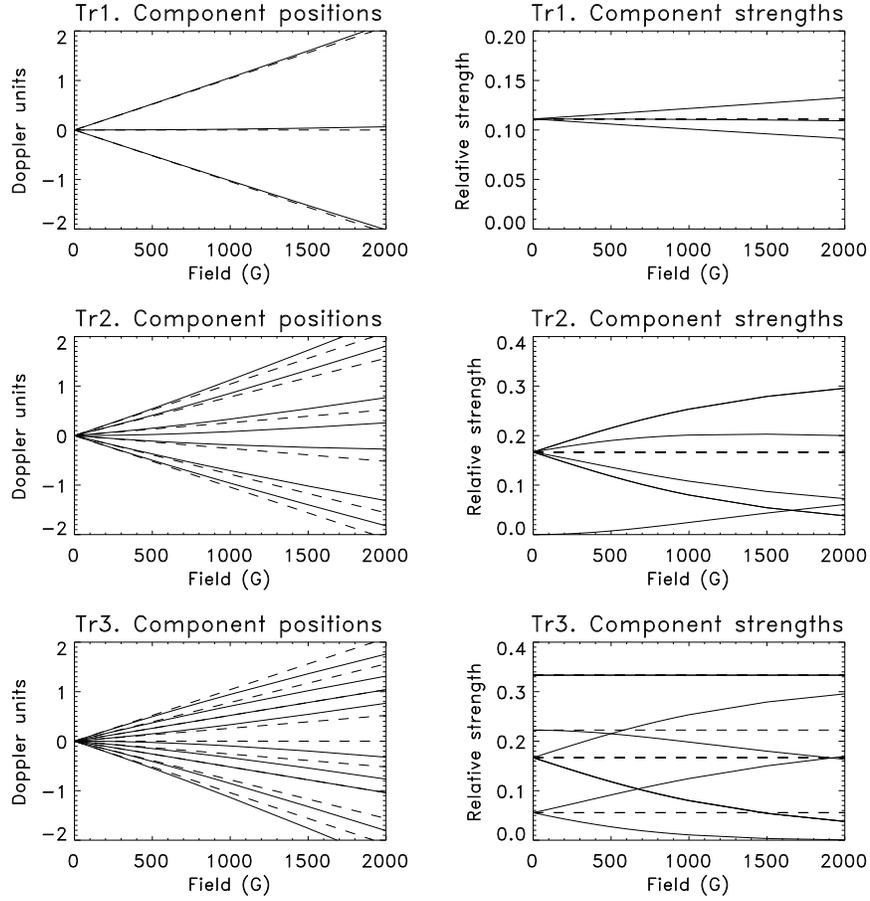}
\caption{Positions and strenghts of the Zeeman components
  as a function of the magnetic field. In all panels the solid lines
  represent calculations considering IPBS while the dashed lines represent
  those using the LZS approximation. Note that, in the general IPBS case, the
  Zeeman components exhibit asymmetric displacements and strengths. Also notice
  that the Zeeman IPBS components have strengths that depend on
  the magnetic field.
\label{components}
}
\epsscale{1.}
\end{figure*}

\begin{deluxetable}{ccc}
  \tablewidth{0pt}
  \tablecaption{Component positions and strengths for Tr1 (500~G)
  \label{TblTr1}}
  \tablehead{Type of & Position & Strength \\
          component &  (Doppler units) &   }
  \startdata  
$\sigma^+$ &     0.524     &        0.106    \\
$\pi$ &   0.004       &        0.111    \\
$\sigma^-$  & -0.516     &        0.116    \\
  \enddata
\end{deluxetable}
\begin{deluxetable}{ccc}
  \tablewidth{0pt}
  \tablecaption{Component positions and strengths for Tr2 (500~G)
  \label{TblTr2}}
  \tablehead{Type of & Position & Strength \\
          component &  (Doppler units) &   }
  \startdata  
$\sigma^+$ &     0.542     &        0.136    \\
$\sigma^+$ &     0.410     &        0.118    \\
$\pi$ &     0.149     &        0.118    \\
$\pi$ &    0.022     &      0.007      \\
$\pi$ &    -0.111     &        0.215    \\
$\sigma^-$  & -0.371     &        0.215    \\
$\sigma^-$  & -0.499     &        0.190    \\
  \enddata
\end{deluxetable}
\begin{deluxetable}{ccc}
  \tablewidth{0pt}
  \tablecaption{Component positions and strengths for Tr3 (500~G)
  \label{TblTr3}}
  \tablehead{Type of & Position & Strength \\
          component &  (Doppler units) &   }
  \startdata  
$\sigma^+$ &     0.495     &       0.091     \\
$\sigma^+$ &     0.371     &        0.215    \\
$\sigma^+$ &     0.260     &        0.333 \\
$\pi$ &     0.111     &        0.215 \\
$\pi$ &   -0.025      &        0.215 \\
$\pi$ &    -0.149     &        0.118 \\
$\sigma^-$  & -0.260  &        0.333 \\
$\sigma^-$  & -0.410  &        0.118 \\
$\sigma^-$  & -0.546  &       0.027 \\
  \enddata
\end{deluxetable}

\subsection{Zeeman components}
\label{comp}

The positions (with respect to the zero-field wavelength) and strengths of
the various $\pi$ and $\sigma$ components of the multiplet are shown in
Fig~\ref{components}. Tables~2 to~4 list the actual IPBS values at 500~G,
which are also useful to match the curves in the left and right panels.
Note that the positions of the Tr1 components are not
significantly disturbed by the presence of the other levels ($J$=1 and $J$=2)
in the field range depicted in the plots. However, the strength of the $\sigma$
components is indeed perturbed, introducing an asymmetry between the left- and
right-handed polarizations. For our discussions in \S\ref{weak} below it is
important to note that this asymmetry increases linearly with the magnetic
field.

The positions of the Tr2 and Tr3 components feel a much stronger
perturbation for fields stronger than $\sim$500~G. This is due to the close
proximity in the energy diagram of the upper levels of these transitions (the
$J=1$ and $J=2$ levels). Similarly to the case of Tr1, the component strengths
of Tr2 and Tr3 become 
asymmetric. The asymmetry is rather linear for weak fields. As the field
increases above $\sim$500~G, however, the behavior of the component strengths
becomes non-linear.

Finally, it is interesting to note the presence of a $\pi$ component in Tr2
that does not exist in the LZS approximation. This new component appears for
non-zero fields in IPBS and becomes more important as the field increases.

The asymmetries between left- and right-handed $\sigma$ components in all three
transitions, are important. They ultimately manifest themselves as asymmetries
in the profiles and lead to net circular polarization even in the absence of
velocity or magnetic field gradients.

\subsection{Weak-field asymptotic behavior}
\label{weak}

Fig~\ref{components} shows that the Zeeman components of the various
transitions tend towards the LZS values as the magnetic strength goes to
zero. One may 
thus expect that the IPBS profiles (plotted in the top row) would become
similar to the LZS profiles in the weak field limit. However, this is not the
case. The profiles obtained from our calculations maintain their shape
unchanged down to the lower limit of our simulation (0.1~G). This apparent
paradox has a simple, albeit slightly counterintuitive, explanation. The
Stokes~V profiles can be described conceptually as the subtraction of two
bell-type functions, similar to Gaussians or Lorentzians. In the weak field
limit the separation between these functions is much smaller than their
width, and the resulting Stokes~V profile is essentially a small residual of
the subtraction. A slight asymmetry between the two functions will show more
conspicuously the closer together they are. Therefore, reducing the field
results in ``amplifying'' the red/blue asymmetries. Let us analyze this in
some detail considering a very simple scenario. Suppose that we have a
Stokes~V profile that is the result of subtracting two Gaussians separated by
a small distance $2\Delta x$:
\begin{equation}
\label{prof}
V(x)=A {\rm e}^{-(x-\Delta x)^2} - B {\rm e}^{-(x+\Delta x)^2} \, .
\end{equation}
The zero-crossing wavelength ($x_0$) of $V(x)$ is given by the condition:
\begin{equation}
A {\rm e}^{-(x_0-\Delta x)^2} = B {\rm e}^{-(x_0+\Delta x)^2} \, ,
\end{equation}
which, taking logarithms on both sides, leads to:
\begin{equation}
x_0={\log(B/A) \over {4 \Delta x}} \, .
\end{equation}
If the asymmetry is small ($B/A \simeq 1$), a Taylor expansion of the
logarithm yields:
\begin{equation}
x_0 \simeq { B - A \over 4 B \Delta x} \, .
\end{equation}
For weak fields, the asymmetry $B-A$ depends linearly on the field ($\Delta
x$). We then conclude that:
\begin{equation}
\label{x0}
x_0 \propto {1 \over 4 B}  \, ,
\end{equation}
i.e., the zero-crossing wavelength is shifted and does {\it not} depend on
the magnetic field. In the derivation above we have been concerned only with
strength asymmetries, neglecting the effect of asymmetric component
positions. If we denote by $\Delta x_A$ and $\Delta x_B$ the (different)
component positions, a similar straightforward reasoning can be conducted to
conclude that:
\begin{equation}
\label{x02}
x_0 \simeq { B-A \over 4 B \overline{\Delta x}} + {1 \over 2}(\Delta x_A
- \Delta x_B) \, ,
\end{equation}
where $\overline {\Delta x}=(\Delta x_A + \Delta x_B)/2$. It is evident from Eq~(\ref{x02}) that position asymmetries have only a
second-order effect on $x_0$, since  $(\Delta x_A - \Delta x_B)$ is
proportional to $\overline {\Delta x}$. We verified this property with our 
simulations. 

Consider now the area asymmetry:
\begin{equation}
\label{area}
{\cal A}={\int_{-\infty}^{\infty} V(x) dx \over \int_{-\infty}^{\infty} |V(x)| dx}
\, .
\end{equation}
In the weak-field limit, the denominator of Eq~(\ref{area}) is directly
proportional to the field. Thus:
\begin{equation}
\label{area2}
{\cal A} \simeq {c \over \Delta x} \int_{-\infty}^{\infty} V(x) dx \, ,
\end{equation}
where $c$ is a constant. Expanding the exponentials in Eq~(\ref{prof}) in a
Taylor series and 
truncating to first order, we obtain that for $\Delta x \ll 1$:
\begin{equation}
 {\rm e}^{-(x-\Delta x)^2} \simeq {\rm e}^{-x^2} +2 x {\rm e}^{-x^2}
 \Delta x \, .
\end{equation}
Thus:
\begin{equation}
\label{vweak}
V(x) \simeq (A-B){\rm e}^{-x^2} + 2x{\rm e}^{-x^2}\Delta x (A+B) \, .
\end{equation}
Inserting this into Eq~(\ref{area2}) and
taking into account the odd parity of the second term in Eq~(\ref{vweak}), it
is evident that:
\begin{equation}
{\cal A} \simeq c \sqrt \pi { A- B \over \Delta x}\, .
\end{equation}
Therefore, the net area asymmetry ${\cal A}$ remains constant even for fields
approaching zero. All of the asymptotic properties of Stokes~V derived in
this section (constancy of the zero-crossing wavelength, net asymmetry,
relative importance of weight and position asymmetries) have been confirmed
by further modeling using our numerical code. It is important to note that
the results we present here do not violate the principle of spectroscopic
stability, since the Zeeman components (and their center of gravity)
approach asymptotically the LZS values.

\begin{figure}
\plotone{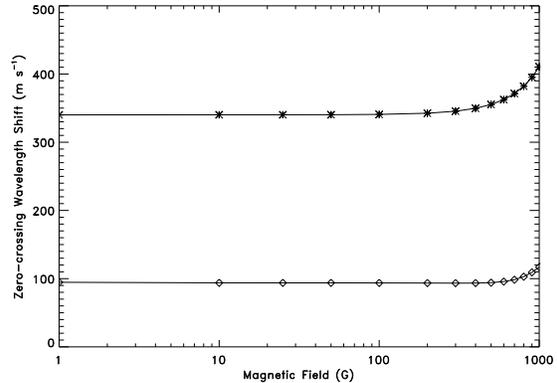}
\caption{Relative Stokes~V zero-crossing shift between IPBS and LZS
  profiles (absolute value), as a function of the field strength. Stars: blue
  component (Tr1). Diamonds: red 
  component (Tr2+Tr3). The Tr1 shift is towards the blue, whereas the Tr2+Tr3
  shift is towards the red.
\label{zerocross}
}
\end{figure}

\begin{figure}
\plotone{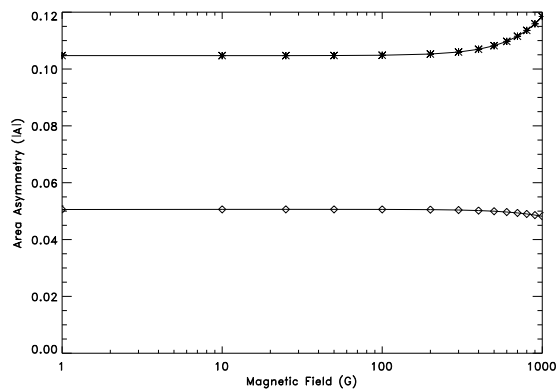}
\caption{Area asymmetry ($|{\cal A}|$) of the IPBS profiles as a function of
  the field strength.  
  Stars: blue component (Tr1). Diamonds: red
  component (Tr2+Tr3). Tr1 has a larger red lobe, whereas Tr2+Tr3 has a
  larger blue lobe.
\label{asym}
}
\end{figure}

The asymptotic values that we
obtained for Tr1 are $x_0$=340~m~s$^{-1}$ and $|{\cal A}|$=0.11 (with the
red lobe having a larger area). It must be noted that
this value of ${\cal A}$ is influenced by the blend with the blue wing of
the red component. We estimated the effect of this blend by calculating ${\cal
  A}$ for the LZS profile (which should be zero), obtaining $|{\cal
  A}_{LZS}|$=0.04. In this case it is the blue lobe that has larger
area. Therefore, the presence of the blend has the effect of reducing the
$|{\cal A}|$ of the IPBS profile. The actual value of $x_0$ is significantly
model-dependent. In particular, it is very sensitive to the Doppler width of
the line. Doubling this parameter in our model, resulted in a factor 4
increase of $x_0$. For the red component of the multiplet (Tr2 and Tr3), we
obtained $x_0$=95~m~s$^{-1}$ and $|{\cal A}|$=0.05. Here, as opposed to Tr1,
the zero-crossing is redshifted and the blue lobe has a larger area. The
dependence of $x_0$ and $|{\cal A}|$ for both lines are plotted in
Figs~\ref{zerocross} and~\ref{asym}.

\begin{figure}
\plotone{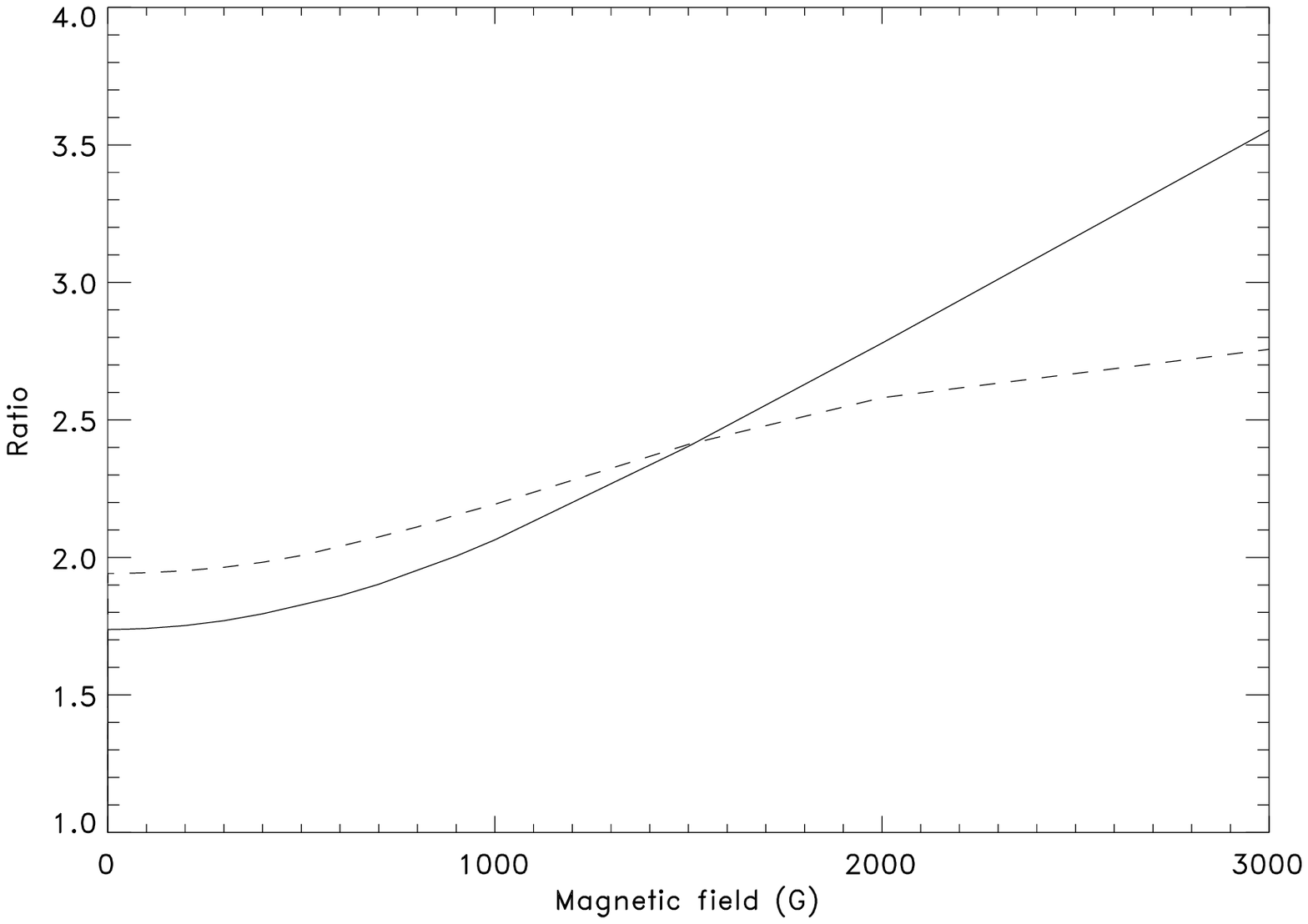}
\caption{Ratio of the Stokes~V amplitudes of the red (Tr2+Tr3) and blue (Tr1)
  components of the multiplet, as a function of the magnetic field for a
  given model. Solid: IPBS. Dashed: LZS.
\label{vratio}
}
\end{figure}

\section{Conclusions}
\label{conclusions}

The calculations reported in this paper show that IPBS effects need to be
taken into 
account for proper diagnostics based on the \ion{He}{1} multiplet
at 10830~\AA . This is a very interesting set of lines, with some unique
properties that make it an ideal tool for the 
exploration of chromospheric and coronal magnetism. Care must be taken, 
however, to consider all the relevant physical processes that intervene in
the formation of the lines.

The (still few) works existing in the literature regarding Stokes
observations of these lines do not seem to agree on such
fundamental parameters as: the rest wavelength of the Tr1 line; the
effective Land\'e factor of the Tr3 line; the relative strengths of the
transitions (compare the values given in \citeNP{RSL95}; \citeNP{LWK+04}; this
work). These are basic parameters that should be well known, either from  
laboratory experiments or from very simple calculations. We believe that, at
least partly, the existing confusion might have its origin in neglecting
IPBS effects. We have shown that IPBS is responsible for a non-Doppler
blueshift of Tr1, systematic profile asymmetries and a reduced amplitude of
Stokes~V in the Tr2+Tr3 blend. As an illustrative example, consider the
simple line-ratio experiment represented in Fig~\ref{vratio}. For a fixed
model atmosphere, the ratio of Stokes~V amplitudes in the red (Tr2+Tr3) and
the blue (Tr1) components of the multiplet depends smoothly on the magnetic
field. It is clear from the figure that neglecting IPBS leads to an erroneous
estimate of the field. Interestingly, the IPBS curve is steeper than the LZS
one, indicating that it is more sensitive to the field.

Some IPBS signatures described in this paper are probably present in published
observations. If one compares the Tr1 Stokes~V profile
depicted in our Fig~\ref{fig:profiles} (upper right panel) with those of
\citeN{LWK+04} (Figs~4 
and~5), the resemblance is striking. More specifically, the IPBS blueshift
and the net offset towards negative values both become immediately
apparent. Profile asymmetries are also present in their observations.
However, it is not possible to assess whether these arise from IPBS or from
velocity or magnetic field gradients.

\acknowledgements
This work has been partially funded by the Spanish
Ministerio de Educaci\'on y Ciencia through project
AYA2001-1649 and by the European Solar Magnetism Network.


\end{document}